\documentclass{emulateapj}
\usepackage{gensymb}
\usepackage{upgreek}
\usepackage{color}
\def\spitz{{\it Spitzer}}
\def\wise{{\it WISE}}

\def\micron{$\upmu$m}

\begin{document}

\shortauthors{Esplin et al.}
\shorttitle{WISE Survey of Disks in Upper Sco}

\title{A WISE Survey of Circumstellar disks in the Upper Scorpius Association\altaffilmark{1}}

\author{
T. L. Esplin\altaffilmark{2,3}, 
K. L. Luhman\altaffilmark{4,5},
E. B. Miller\altaffilmark{4}, and
E. E. Mamajek\altaffilmark{6,7}}

\altaffiltext{1}{Based on observations made with the {\it Wide-field Infrared 
Survey Explorer}, the Two Micron All Sky Survey, the {\it Gaia} mission,
the United Kingdom Infrared Telescope Infrared Deep Sky Survey,
the NASA Infrared Telescope Facility,
the Cerro Tololo Inter-American Paranal Observatory, 
the Southern Astrophysical Research Telescope,
and the {\it Spitzer Space Telescope}.}
\altaffiltext{2}{Steward Observatory, University of Arizona, Tucson, AZ, 85719, 
USA; taranesplin@email.arizona.edu}
\altaffiltext{3}{Strittmatter Fellow}
\altaffiltext{4}{Department of Astronomy and Astrophysics, The Pennsylvania
State University, University Park, PA 16802, USA}
\altaffiltext{5}{Center for Exoplanets and Habitable Worlds,
The Pennsylvania State University, University Park, PA 16802, USA}
\altaffiltext{6}{Jet Propulsion Laboratory, California Institute of Technology,
4800 Oak Grove Dr., Pasadena, CA 91109, USA}
\altaffiltext{7}{Department of Physics and Astronomy, University of Rochester,
500 Wilson Blvd., Rochester, NY 14627, USA}

\begin{abstract}

We have performed a survey for new members of the Upper Sco association
that have circumstellar disks using mid-infrared photometry from
the {\it Wide-field Infrared Survey Explorer (WISE)}.
Through optical and near-infrared spectroscopy, we have confirmed 185
candidates as likely members of Upper Sco with spectral types ranging
from mid-K to M9. They comprise $\sim$36\% of the known disk-bearing
members of the association.
We also have compiled all available mid-infrared photometry from
{\it WISE} and the {\it Spitzer Space Telescope} for the known members
of Upper Sco, resulting in a catalog of data for 1608 objects.
We have used these data to identify the members that exhibit excess emission
from disks and we have classified the evolutionary stages of those disks
with criteria similar to those applied in our previous studies of Taurus and
Upper Sco.
Among 484 members with excesses in at least one band (excluding 
five Be stars),
we classify 296 disks as full, 66 as evolved, 19 as transitional,
22 as evolved or transitional, and 81 as evolved transitional or
debris. Many of these disks have not been previously reported,
including 129 full disks and 50 disks that are at more advanced
evolutionary stages.

\end{abstract}


\section{Introduction}

Thorough surveys for the disk-bearing members of nearby
star-forming regions and young associations are important for
studies of the evolution of circumstellar disks.
The Upper Scorpius subgroup \citep[$\sim$11~Myr;][]{pec12} in the
Scorpius-Centaurus OB association is a particularly valuable laboratory
for studying disk evolution.
It contains one of the nearest and richest populations of disk-bearing
stars \citep[$d\sim145$~pc, $N_{\rm disk}\sim500$,][this work]{pre08,luh12}
and those disks span a wide range evolutionary stages \citep{car09}.
In addition, because the natal cloud has dispersed, the members have
relatively low extinction ($A_V\lesssim3$), enabling observations of ultraviolet
and optical signatures of accretion.

Most of the known circumstellar disks in Upper Sco have been identified
using mid-infrared (IR) photometry from the {\it Spitzer Space
Telescope} \citep{wer04} and the {\it Wide-field Infrared Survey Explorer} 
\citep[{\it WISE};][]{wri10,rie05,car06,car08,car09,che05,che11,ria09,ria12,luh12,riz12,riz15,daw13,pec16}.
However, no study has searched all known members for evidence of disks
in a uniform manner. In addition, the current membership list for Upper Sco
is likely to be incomplete, so the same is likely true for the disk-bearing
members as well.
To work towards a more complete census of disks in Upper Sco,
we have performed a survey for new disk-bearing members using data
from {\it WISE} and wide-field optical and near-IR surveys
(Section~\ref{sec:search}) and we have classified the disks
among all known members with the methods applied to an earlier compilation 
of members by \citet{luh12} (Section~\ref{sec:class}).

\section{Identification of New Disk-bearing Members of Upper Sco}
\label{sec:search}

\subsection{Boundary between Upper Sco and Ophiuchus}
\label{sec:oph}

\begin{figure}[h]
	\centering
	\includegraphics[trim = 0mm 0mm 0mm 0mm, clip=true, scale=.8]{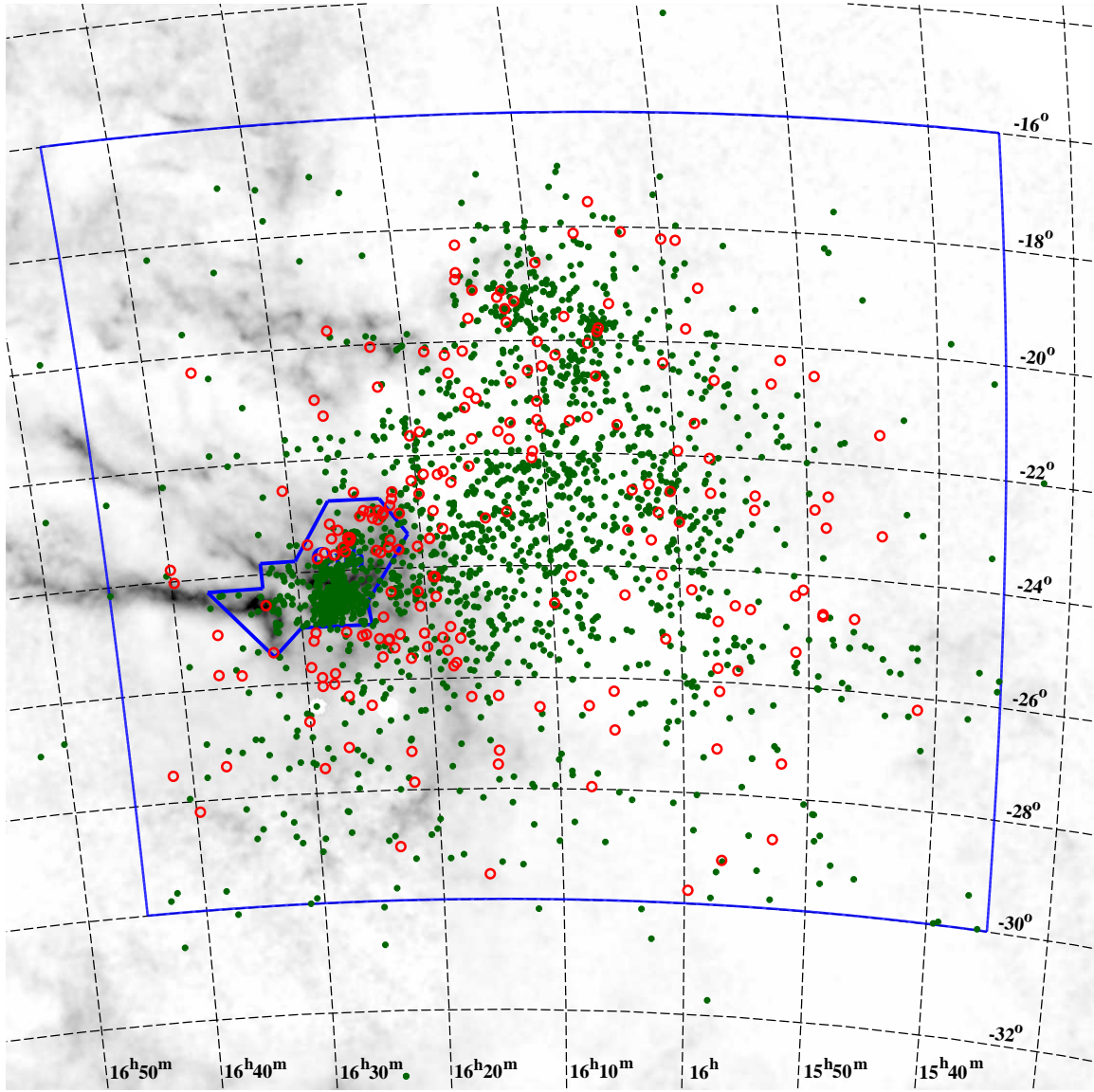}
\caption{
Spatial distribution of the previously known young stars in the Upper Sco 
association and Ophiuchus star-forming region (filled circles) and new stars
with disks that we have found (open circles).  
In our survey, we have considered the area from
$\alpha=15^{\rm h}35^{\rm m}$ to $16^{\rm h}45^{\rm m}$ and
$\delta=-30$ to $-16\arcdeg$ (rectangle). Our adopted boundary between
Ophiuchus and Upper Sco is indicated (small polygon, Table~\ref{tab:oph}).
The Ophiuchus dark clouds and the diffuse clouds across portions of Upper
Sco are displayed with a map of extinction \citep[gray scale,][]{dob05}.
}
\label{fig:spatial}
\end{figure}

The Ophiuchus star-forming region \cite[$\sim$1~Myr,][]{wil08} is projected 
against the eastern edge of Upper Sco, as illustrated in 
Figure~\ref{fig:spatial}. 
To help define our survey field for Upper Sco, we describe our
selection of a boundary between the two populations.

\begin{figure}[h]
	\centering
	\includegraphics[trim = 0mm 0mm 0mm 0mm, clip=true, scale=.9]{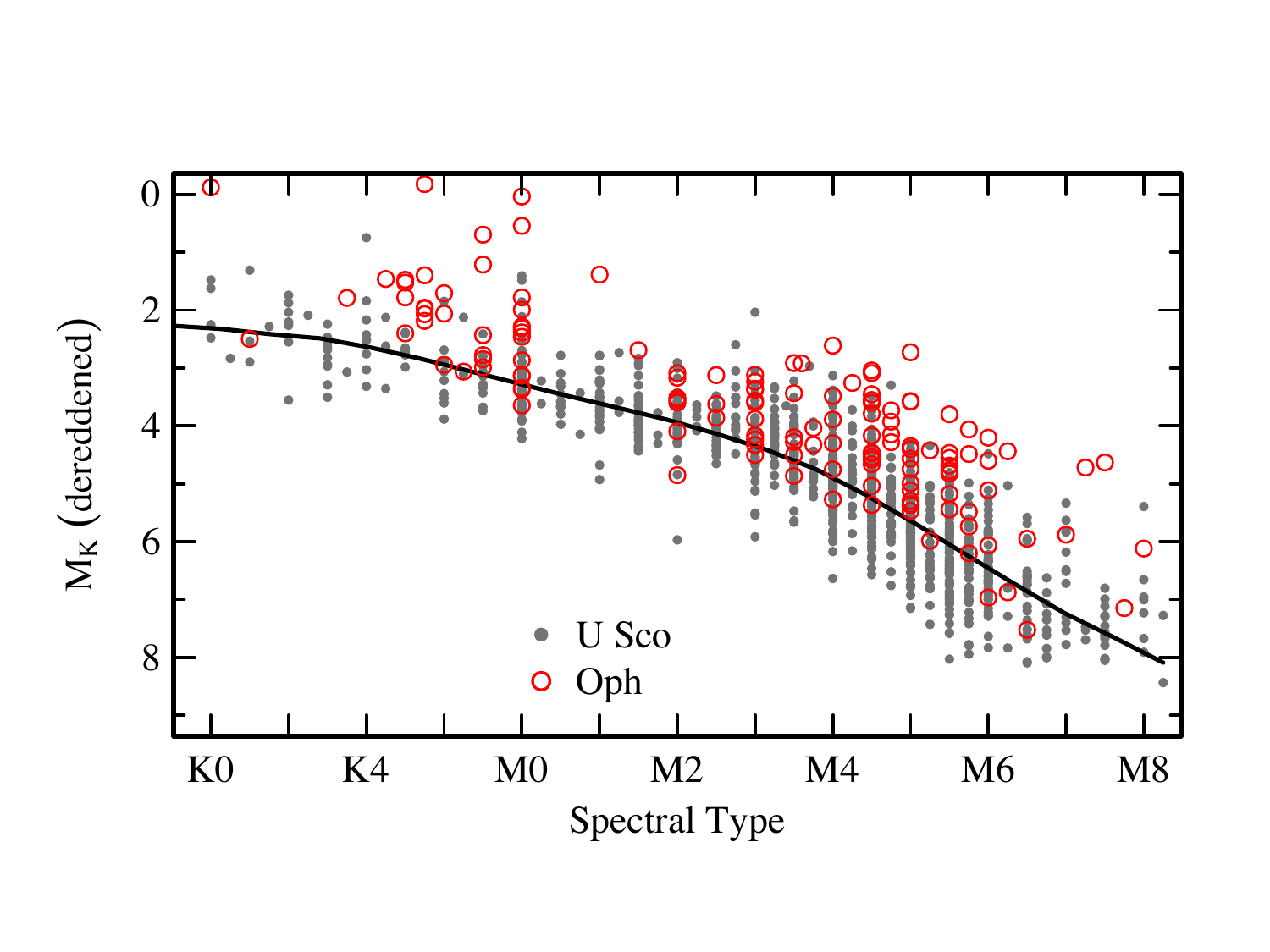}
\caption{
Extinction-corrected $M_K$ as a function of spectral type for K0--M8 members
of Upper Sco \citep[filled circles,][references therein]{luh18} and the cloud 
core of Ophiuchus \citep[open circles,][references therein]{wil08} that have parallaxes 
from Gaia DR2.
The median of the sequence for Upper Sco is indicated (solid line).}
\label{fig:svk}
\end{figure}

We have compiled a list of all stars that have been identified
as likely members of either Upper Sco or Ophiuchus 
\citep[][references therein]{wil08,luh18}.
We adopted near-IR photometry for those stars from the Point Source
Catalog of the Two Micron All Sky Survey \citep[2MASS;][]{skr06}
and the tenth data release of the United Kingdom Infrared 
Telescope Infrared Deep Sky Survey \citep[UKIDSS;][]{law07}.
When both catalogs contained photometry for a given star, we adopted the
measurement with the smaller error.
We estimated the extinction for each object using $J-H$ (or $H-K_s$
when $J-H$ was unavailable) and the intrinsic
photospheric values from \citet{luh10} and  \citet{pec13}.
We calculated extinction-corrected absolute magnitudes in $K_s$ ($M_K$) for
the stars that have parallax measurements with errors of $\leq$10\% from the
second data release (DR2) of {\it Gaia} \citep{gai16,gai18}.
In Figure~\ref{fig:svk}, we plot extinction-corrected $M_K$
as a function of spectral type for the members of Upper Sco from
\citet{luh18} and the  Ophiuchus members projected against the L1688 cloud.
The members of the latter are systematically brighter at a given spectral type,
which is a reflection of its younger age.

\begin{figure}[h]
	\centering
	\includegraphics[trim = 0mm 0mm 0mm 0mm, clip=true, scale=.6]{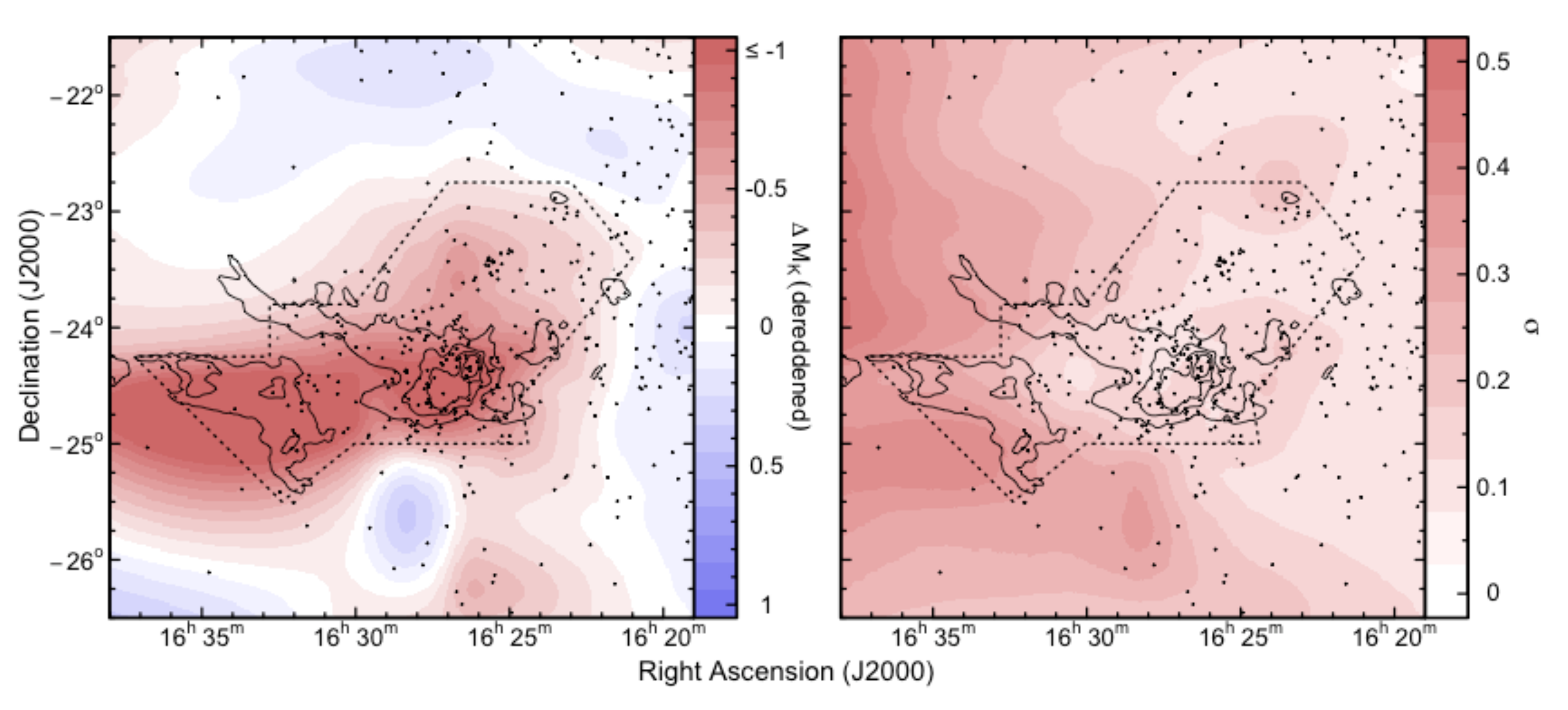}
\caption{
Fit to the offsets in extinction-corrected $M_K$ from the median sequence for
K0--M8 members of Upper Sco in Figure \ref{fig:svk} as a function of location 
(left) and the corresponding error value of $\Delta M_K$ (right).
The Ophiuchus clouds are represented by a map of $^{13}$CO 
emission \citep[contours;][]{rid06}.  We adopted a boundary for Ophiuchus
that encompasses the dark clouds and regions in which the values of 
$\Delta M_K$ are low (i.e., brighter) and their errors are low.}
\label{fig:age}
\end{figure}

We have used the brighter values of  $M_K$ for members of Ophiuchus
to define a spatial boundary between Ophiuchus and Upper Sco.
We computed the median values of extinction-corrected  $M_K$ as a function
of spectral type for Upper Sco by applying local linear quantile regression 
with the function {\tt lprq} in the R package {\it quantreg} \citep{koe16}. 
For each young star in the vicinity of Ophiuchus, we calculated the offset
in extinction-corrected $M_K$ from the median sequence of Upper Sco
($\Delta M_K$). Negative values of $\Delta M_K$ correspond to positions
above (brighter than) the sequence for Upper Sco.
We derived the average $\Delta M_K$ as a function of spatial
position using a single order local regression (LOESS) as implemented in R 
\citep{Rcore13}, which we show in Figure~\ref{fig:age}.
We also include in Figure~\ref{fig:age} estimates of the errors in those
average values of $\Delta M_K$, which are based on bootstrapping. 
The Ophiuchus cloud core as well as large areas to the east and north contain
stars that are significantly brighter,  and hence younger, 
than the median sequence for Upper Sco.
We have selected a boundary for Ophiuchus that encompasses the dark clouds
and the areas in which $\Delta M_K$ and the corresponding errors are low.
The resulting boundary is defined in Table~\ref{tab:oph} and is marked
in Figures~\ref{fig:spatial} and \ref{fig:age}.
We use this boundary for separating young stars into membership lists
for Upper Sco and Ophiuchus in our survey and in \citet{luh18}.
However, a single spatial boundary is unlikely to be sufficient to fully
separate the two populations.
In fact, multiple populations are likely present within the region that
we have defined for Ophiuchus \citep{bou92,mar98,wil05}.
For instance, \cite{pil16} have identified a cluster of stars surrounding
the star $\rho$~Oph that is noticeably older than the stars associated
with the cloud core (which is south of $\rho$~Oph).

\subsection{Candidate Members from WISE}
\label{sec:subsearch}

Our survey for new disk-bearing members of Upper Sco is based primarily
on data from the \wise\ mission \citep{wri10}.
During its fully cryogenic phase between January and July of 2010, \wise\
obtained images of the entire sky in four bands centered at 3.4, 4.6, 12,
and 22~\micron, which are denoted as $W1$ through $W4$.
The telescope continued to collect data in $W1$, $W2$, and $W3$ through
September of 2010 with cryogens remaining in one of its two tanks.
Following exhaustion of the second tank, \wise\ operated in
$W1$ and $W2$ through January of 2011 \citep[{\it NEOWISE},][]{mai11}.
The images exhibited an angular resolution of $\sim6\arcsec$ and
$\sim12\arcsec$ for $W1$ through $W3$ and $W4$, respectively. 
Data from the fully cryogenic phase and the entire 13-month mission
were released in the \wise\ All-Sky Source Catalog and the AllWISE Source
Catalog, respectively.
We initially used the {\it WISE} All-Sky Catalog for our survey.
We repeated our analysis with the AllWISE Source Catalog when it was released.

We began our search by retrieving all \wise\ sources
between right ascensions ($\alpha$) of 15$^{\rm h}$35$^{\rm m}$ and
16$^{\rm h}$45$^{\rm m}$ and declinations ($\delta$) of $-30\arcdeg$ and
$-16\arcdeg$ (J2000), which encompass most of the known members of Upper Sco
(Figure~\ref{fig:spatial}). 
For our survey of Upper Sco, we focused on \wise\ sources that are outside of
the boundary of Ophiuchus that was adopted in the previous section.
In addition, we have considered only sources that have errors $\leq0.1$ in $W1$ 
and $W2$ and are not flagged as a diffraction spike in either of those bands by the 
parameter {\tt cc\_flag}. Measurements in $W3$ and $W4$ with errors $>0.1$
were not used.

\begin{figure}[h]
	\centering
	\includegraphics[trim = 0mm 0mm 0mm 0mm, clip=true, scale=1]{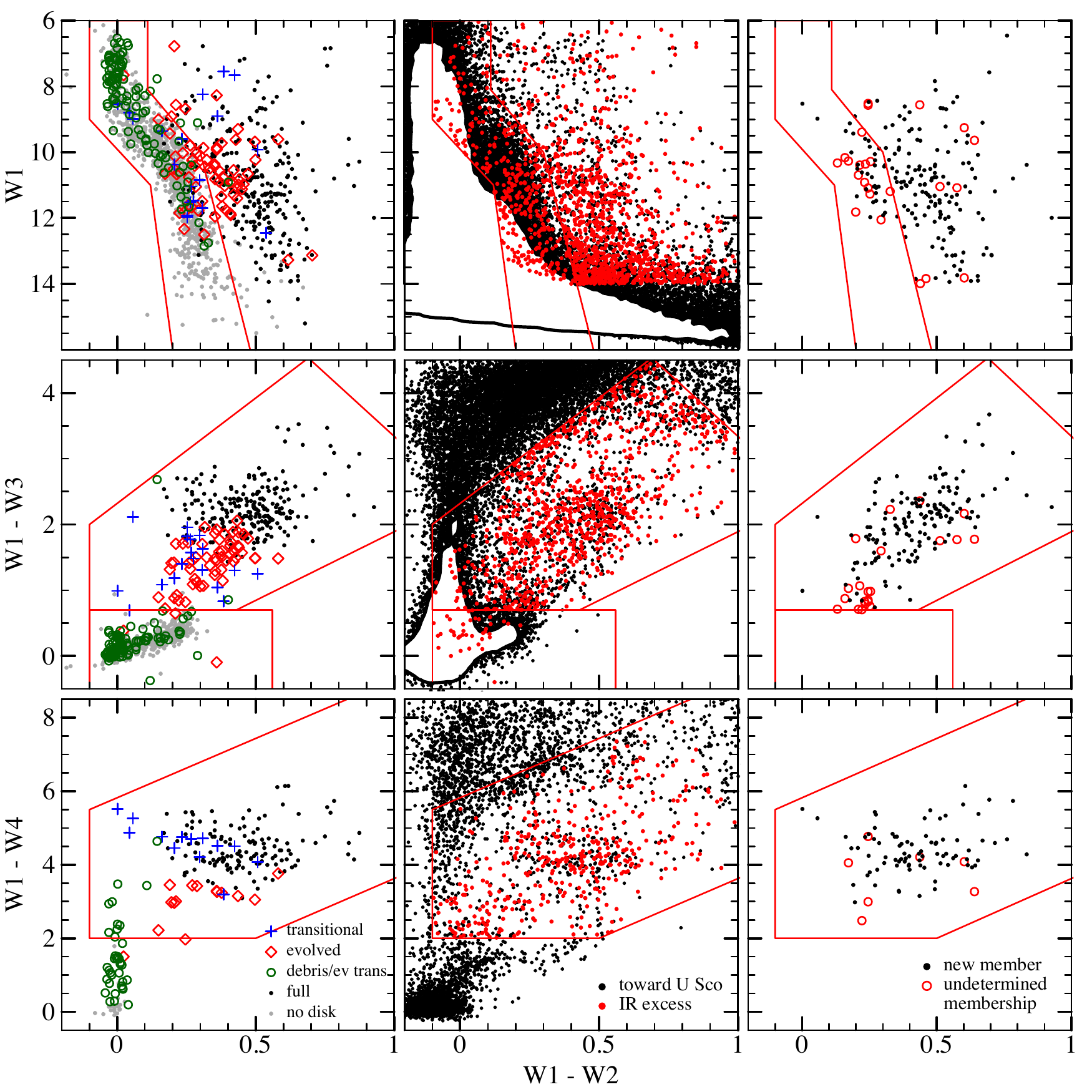}
\caption{
Left: {\it WISE} CMD and color-color diagrams for the known members of Upper
Sco. We have defined regions that separate most disk-bearing members from
those without disks (red lines). We have indicated full disks
(black points),
evolved disks (crosses), transitional disks (pluses),
debris disks or evolved transitional disks (open circles),
and other members of Upper Sco (gray points).
Middle: Among AllWISE sources projected against Upper Sco that are not
known members (filled circles), we have selected candidate disk-bearing members
using the boundaries defined on the left (red filled circles).
Right: Candidates that have been confirmed as members through spectroscopy 
(filled circles, Table~\ref{tab:specu}) and candidates that lack spectroscopy 
(open circles, Table~\ref{tab:cand})). 
}
\label{fig:wise}
\end{figure}

We have constructed a color-magnitude diagram (CMD) and two color-color
diagrams from the \wise\ data for the known members of Upper Sco, which are
shown in the left column of Figure~\ref{fig:wise}. The members that have disks
are indicated (Section~\ref{sec:term}).
In each diagram, the members without excesses form a sequence or clump with
bluer colors while members with excesses have a broader distribution of 
redder colors. 
We defined boundaries in those diagrams that separate 
members with and without excesses. Although colors of $W1-W4\gtrsim0.5$
indicate an excess relative to photospheric values, we adopted a boundary
of $W1-W4=2$ because of the large number of sources at $W1-W4\sim0.5$--2,
nearly all of which are probably nonmembers. As a result, our survey is not
sensitive to disk-bearing stars with small excesses in $W4$.
We note that WISEA J161935.70$-$195043.0 \citep[EPIC 205008727,][]{cod17}
is the reddest known member in the \wise\ colors with $W1-W2=1.3$ and
$W1-W4=6.7$, placing it beyond the limits of the diagrams in
Figure~\ref{fig:wise}.
Its colors are indicative of a protostar or a star with an edge-on disk.
The presence of the former in Upper Sco would be surprising given the
age of the association. The optical spectrum from \citet{cod17}
exhibited strong continuum veiling and emission lines and a spectral type
of K7--M3. Our near-IR spectrum of the star shows similar characteristics.

In the middle column of Figure~\ref{fig:wise}, we show the CMD and color-color
diagrams for \wise\ sources toward Upper Sco that are not known members
of the association. 
We have marked sources that exhibit an IR excess in at least one
diagram (based on the boundaries that we have defined), have excesses
in all available bands longward of the shortest wavelength at which an excess
is present, and have $W1\leq14$.
Sources that lack detections in bands at longer wavelengths are included as 
candidates only if those limits are consistent with an excess.
These criteria produced $\sim$800 candidate disk-bearing members of Upper Sco.

\subsection{Additional Membership Constraints}

In a large survey field like the one that we have considered in Upper Sco,
candidate disk-bearing objects identified with \wise\ photometry can contain 
significant contamination from field stars and galaxies \citep{esp14}.
As a result, we have attempted to further refine our sample of 
\wise\ candidates by applying additional membership constraints.

We visually inspected the images of the candidates from \wise, 2MASS,
and the Digitized Sky Survey.
We rejected candidates that moved noticeably between different epochs (i.e.,
those with high proper motions), appeared to be galaxies based on
their extended emission, or were not reliably detected or were blended
with other objects in the \wise\ bands that seemed to show excesses.
We also removed sources that have been classified as field stars or galaxies
through spectroscopy from previous studies.

\begin{figure}[h]
	\centering
	\includegraphics[trim = 0mm 0mm 0mm 0mm, clip=true, scale=1]{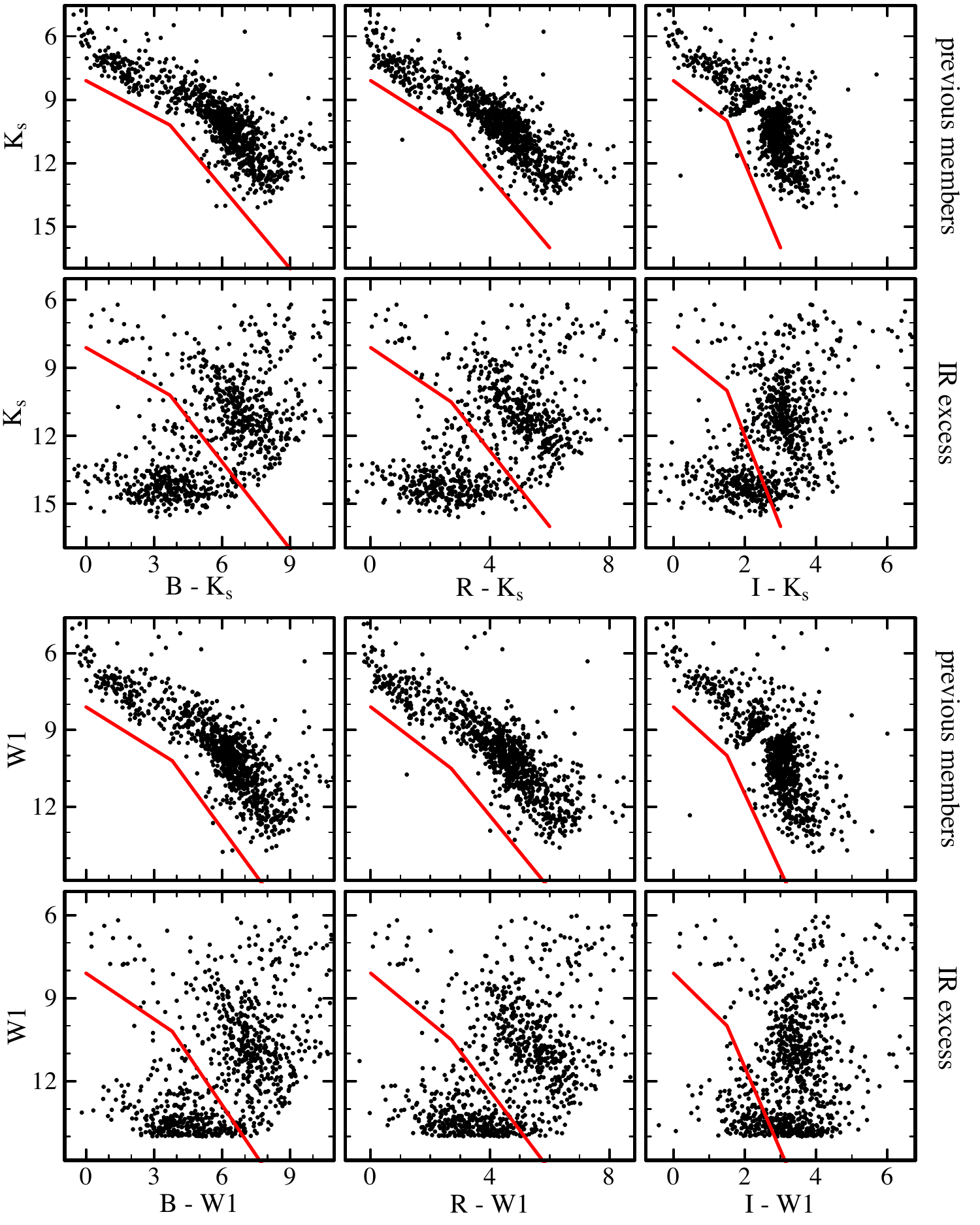}
\caption{
CMDs of the previously known members of Upper Sco and candidate disk-bearing
stars from Figure \ref{fig:wise}. 
These diagrams were constructed with data from {\it WISE} ($W1$),
2MASS ($K_s$), and USNO-B1.0 ($BRI$). 
Candidates that appear below the solid boundaries are rejected as nonmembers.
}
\label{fig:usno}
\end{figure}

To check whether the \wise\ candidates have optical and near-IR data that
are consistent with membership in Upper Sco, we have plotted in
Figure~\ref{fig:usno} CMDs for
the previously known members and the candidates using $W1$, $BRI$ from the
U.S. Naval Observatory B1.0 Catalog \citep[USNO-B1.0;][]{mon03}, 
and $K_s$ from the 2MASS Point Source Catalog.
For sources with detections in both USNO epochs, we have adopted the more
recent measurements.
In each diagram, we defined a boundary that follows the lower envelope of
the sequence of known members, as indicated in Figure~\ref{fig:usno}. 
Candidates that appear below any of those boundaries are rejected.
We note that CMDs in additional optical and near-IR bands can be constructed
for Upper Sco with more recent wide-field surveys like
UKIDSS, {\it Gaia} \citep{per01}, and Pan-STARRS1 \citep{kai02,kai10}.

Some of the \wise\ candidates appear within the {\it Spitzer} images that
have been obtained for small portions of Upper Sco.
We retrieved the available photometry for those candidates from the 
{\it Spitzer} Enhanced Imaging Products (SEIP) Source List.
We checked whether the {\it Spitzer} data show excesses indicative of disks
for the \wise\ candidates that have excesses in $W2$ and that lack detections
in $W3$ and $W4$. Known members of Upper Sco with excesses at $W2$ 
have colors between the 5.8 and 8.0~\micron\ bands of \spitz\ that 
are greater than 0.3.
Therefore, we rejected the \wise\ candidates
that are bluer than that threshold.

Finally, we checked whether our candidates from \wise\ have proper motions
that are consistent with membership using data from the
fourth and fifth releases of the U.S. Naval Observatory CCD Astrograph Catalog
\citep[UCAC4,UCAC5,][]{zac13,zac17}.
Candidates are retained if the 1$\sigma$ errors of their motions 
overlap with a radius of 10~mas~yr$^{-1}$ from the median value of the
motions of known members. 
This criterion is satisfied by $\sim$90\% of the known members.
A more detailed description of the UCAC5 motions
for the known members of Upper Sco is provided by \citet{luh18}.

The criteria in this section resulted in the rejection of $\sim60$\% of the 
$\sim800$ disk-bearing stars identified with \wise.

\subsection{Spectroscopy of Candidate Members}
\label{sec:new}

\subsubsection{Observations}

We performed spectroscopy on 233 candidate disk-bearing stars 
to measure their spectral types and check for signatures of youth.
Two of these candidates are components of a $1\farcs8$ pair that is
unresolved in the \wise\ data (WISEA J155101.25$-$252310.0).
Twelve candidates in our spectroscopic sample were initially identified
with the {\it WISE} All-Sky Catalog and UCAC4 but are no longer candidates
in our final analysis that uses the AllWISE Source Catalog and UCAC5.
Ten of those stars lack mid-IR excesses or do not satisfy our proper motion
criteria when using the latter catalogs. The remaining stars,
WISE J160027.15$-$223850.5 and WISE J160414.16$-$212915.5,
were absent from the AllWISE Source Catalog.
Although our survey focuses on the area outside of our adopted boundary 
for Ophiuchus, we have included in our spectroscopic sample 29 candidates 
that appear within that boundary.

The spectra were taken with 
SpeX \citep{ray03} at the NASA Infrared Telescope Facility (IRTF), 
the Goodman High Throughput Spectrograph at the Southern Astrophysical 
Research Telescope (SOAR), and the Cerro Tololo Ohio State Multi-Object
Spectrograph (COSMOS) and the Astronomy Research using the Cornell Infra Red
Imaging Spectrograph (ARCoIRIS) at the 4~m Blanco telescope at the Cerro 
Tololo Inter-American Observatory (CTIO).
The instrument configurations are summarized in Table~\ref{tab:log}.
The dates and instruments are indicated for the targets in Upper Sco
and Ophiuchus in Tables~\ref{tab:specu} and \ref{tab:speco}, respectively.

\begin{figure}[h]
	\centering
	\includegraphics[trim = 0mm 0mm 0mm 0mm, clip=true, scale=0.8]{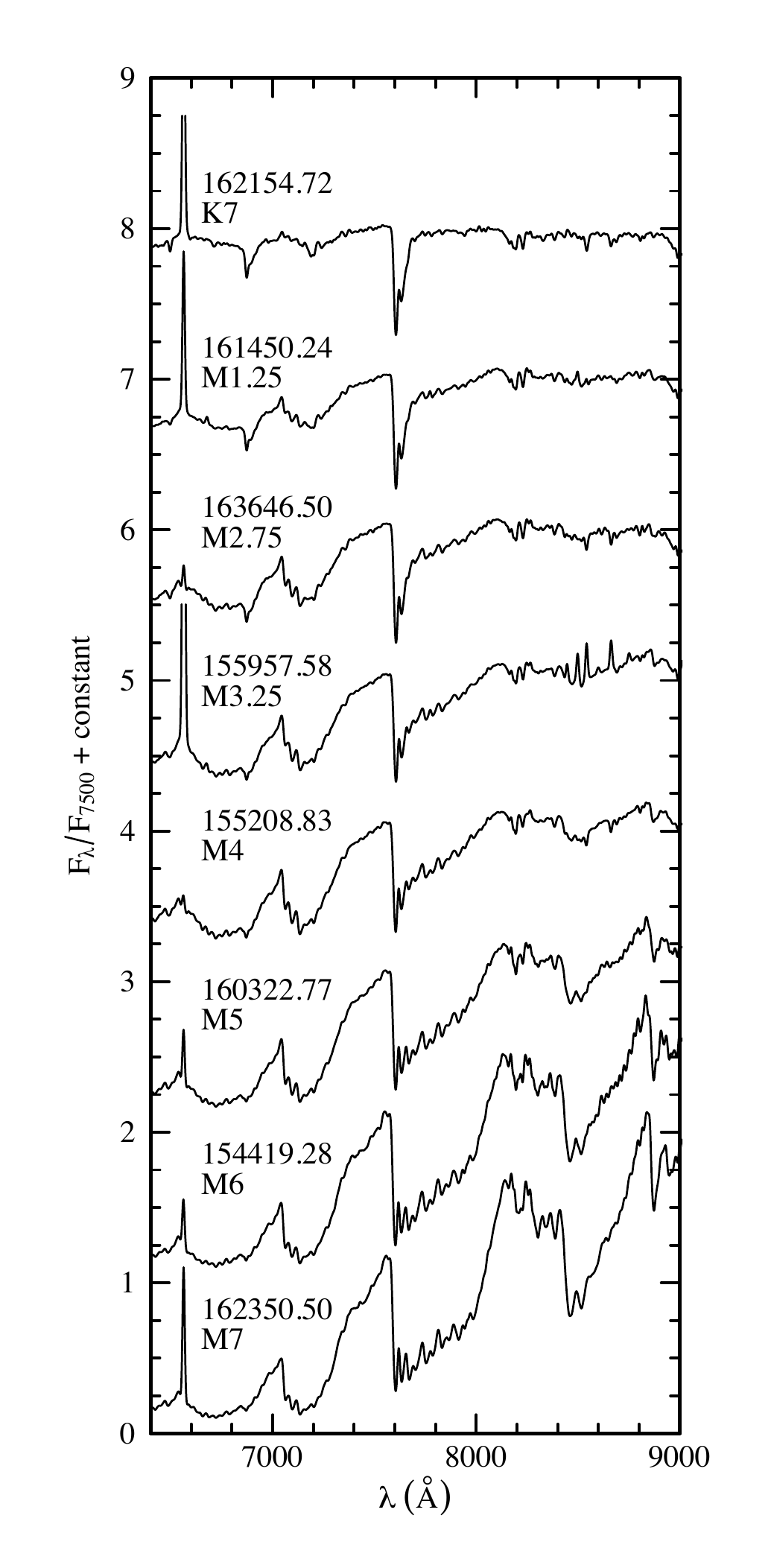}
\caption{
Optical spectra of a sample of new disk-bearing stars (Tables~\ref{tab:specu}
and \ref{tab:speco}), which span the range of spectral types observed. 
These data are displayed at a resolution of 13~\AA.
The data used to create this figure are available.
}
\label{fig:op}
\end{figure}

\begin{figure}[h]
	\centering
	\includegraphics[trim = 0mm 0mm 0mm 0mm, clip=true, scale=0.8]{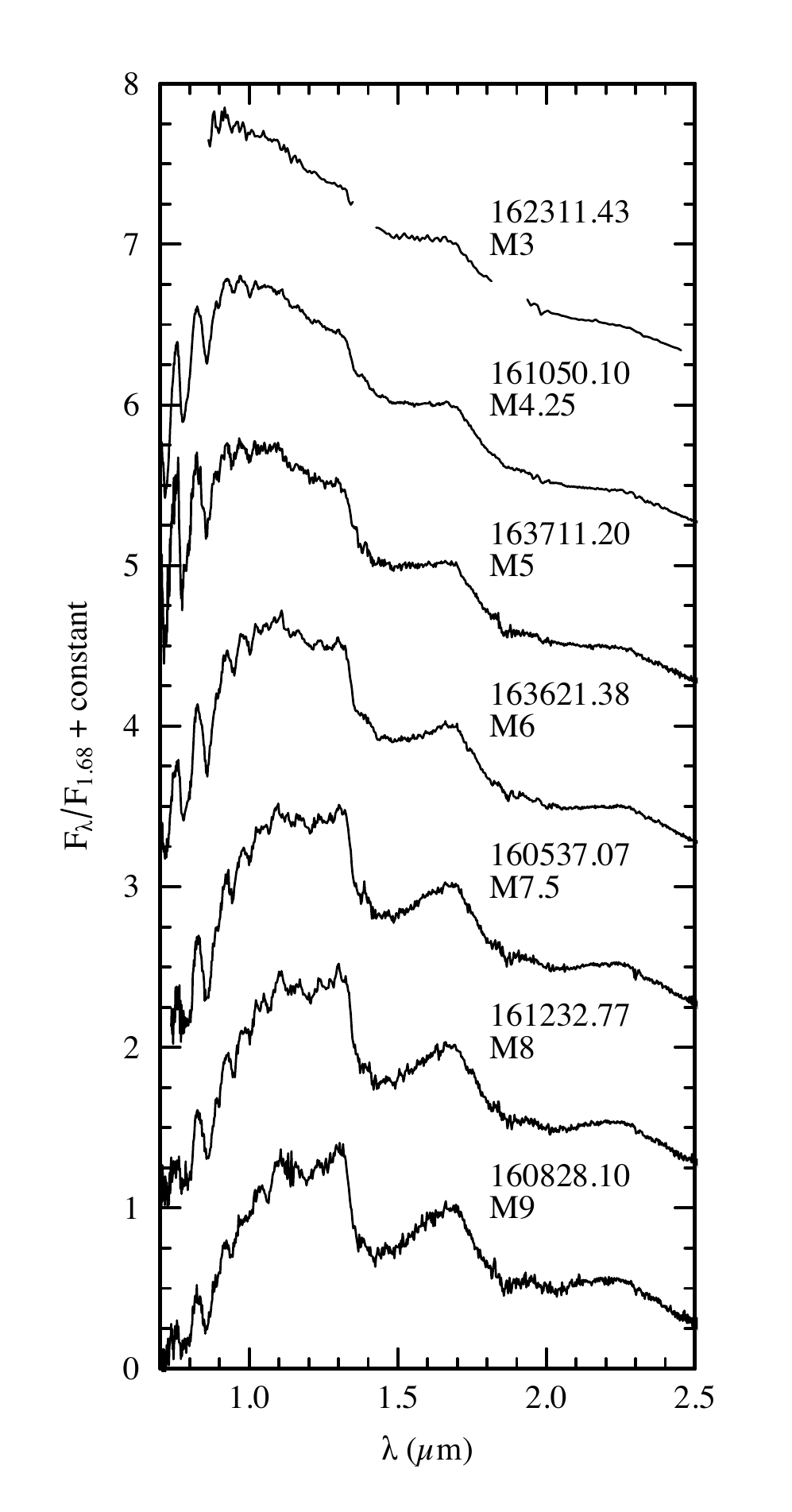}
\caption{
Near-IR spectra of a sample of new disk-bearing stars (Tables~\ref{tab:specu}
and \ref{tab:speco}), which span the range of spectral types observed. 
They have been dereddened to match the slopes of the young standards from
\citet{luh17}. The data used to create this figure are available.
}
\label{fig:ir}
\end{figure}

We reduced the Goodman and COSMOS spectra using routines in IRAF.
The steps of reduction included flat-field correction, the extraction of
spectra, and wavelength calibration. 
The SpeX data were reduced using the Spextool package \citep{cus04} and were
corrected for telluric absorption in the manner described by \cite{vac03}.
We reduced the ARCoIRIS data in the same manner as the SpeX data
using a modified version of Spextool. 
Samples of the reduced optical and near-IR spectra are included in 
Figures~\ref{fig:op} and \ref{fig:ir}, respectively.
All of the reduced spectra from our survey are provided in electronic
files that accompany those figures.

\subsubsection{Spectral Classification}

\begin{figure}[h]
	\centering
	\includegraphics[trim = 0mm 40mm 0mm 110mm, clip=true, scale=0.8]{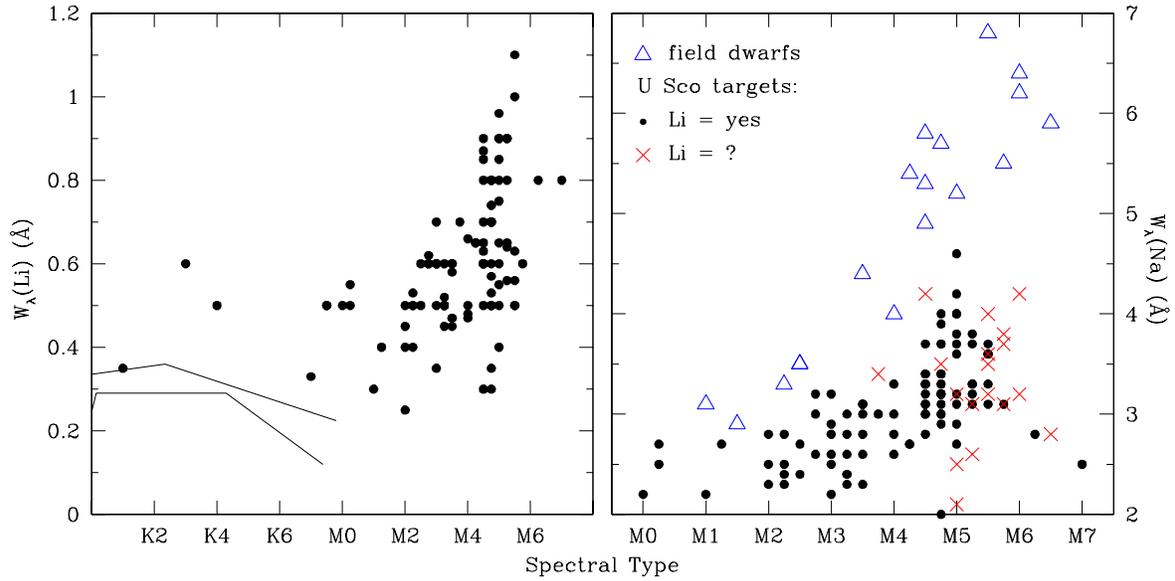}
\caption{
Equivalent widths of Li and Na versus spectral type for candidate disk-bearing
stars observed with optical spectroscopy, except for those classified as giants
(Tables~\ref{tab:specu} and \ref{tab:speco}).
The left diagram contains the detections of Li (filled circles).
In the right diagram, we show Na measurements for those Li-bearing stars
as well as sources that lack useful constraints on Li (crosses).
For comparison, we include the upper envelopes for Li data in IC~2602 (45~Myr)
and the Pleiades (125~Myr) \citep[upper and lower solid lines,][]{neu97}
in the left diagram and Na measurements for a sample of field dwarfs (open
triangles) in the right diagram.
All of the Li and Na data for our targets are consistent with the youth
expected for disk-bearing stars.
}
\label{fig:lina}
\end{figure}

To determine whether the candidates in our spectroscopic sample are likely to
be members of Upper Sco and Ophiuchus, we have measured their spectral types 
and checked their spectra for evidence of youth.
Given the magnitude range of our candidates, 
they should have spectral types of K--M if they are members.
Thus, any targets with earlier types are probably field stars.
Among K and M types, we use Li~I absorption and gravity-sensitive features 
\citep[e.g., Na I, H$_2$O;][]{mar96,luh97,luc01}
to distinguish young objects from field stars.
In Figure~\ref{fig:lina}, we show the Li and Na equivalent widths for
K--M and M types, respectively, for our targets that were observed with optical
spectroscopy with the exception of the stars classified as giants.
We also include the upper envelopes for Li data in IC~2602 (45~Myr) and the
Pleiades (125~Myr) from \citet{neu97} and measurements of Na for a
sample of standard field dwarfs from our previous surveys and \citet{fil16}.
All of the Li and Na data for our targets in Figure~\ref{fig:lina} are
consistent with the youth implied by the IR excesses.

Dwarfs and giants were classified with optical and IR spectra of standards
\citep{hen94,kir91,kir97,cus05,ray09}. We also applied those standard
spectra to optical and IR spectra of young stars at $<$M5 and $<$M0,
respectively.
We classified the optical spectra of young stars at $\geq$M5 with
the average spectra of standard dwarfs and giants \citep{luh97,luh98,luh99}
and we classified the IR spectra of young M-type stars with
standard spectra based on optically-classified young stars \citep{luh17}.

In Tables~\ref{tab:specu} and \ref{tab:speco}, we list the spectral types, 
membership classifications, and equivalent widths of Li and Na (when available)
for our spectroscopic sample.
We have classified 216 of the 233 candidates as young stars. Two of the 
young stars, WISEA J160104.73$-$261653.5 and J162847.63$-$262002.6, are
not considered members of Upper Sco because their proper motions are
inconsistent with membership \citep{tia17}. The remaining 214 young stars
are adopted as members of Upper Sco (185) or Ophiuchus (29).
Some (19) of these members have been independently uncovered by
previous studies \citep{riz15,ans16,pec16,cod17,bes17}.
Those stars can be identified in the compilation of known members of Upper Sco
from \cite{luh18} via the presence of spectral types from both this work and
a previous study.

There remain 30 candidate disk-bearing stars in our survey field for Upper
Sco that lack spectroscopy and that are not rejected by the criteria
for CMDs and proper motions from \citet{luh18}. These stars are listed
in Table~\ref{tab:cand} and are plotted in the CMD and color-color diagrams
in the right column of Figure~\ref{fig:wise}. We also include in those diagrams
the new members of Upper Sco from this work.

\subsection{Completeness of Census of Members with Disks}

In the absence of severe crowding, the AllWISE Source Catalog is
$\sim95$\% complete at $W1=16.9$, $W2=15.5$, $W3=11.6$, and $W4=7.7$
\citep{cut12,cut13}. Meanwhile, our selection criteria include thresholds of
$W1<14$, $W1-W2>0.1$, $W1-W3>0.75$, and $W1-W4>2$.
By combining the completeness limits of AllWISE with our criteria, we estimate
that our sample of candidates should have $\gtrsim95$\% completeness
for stars with excesses in $W2$, $W3$, and $W4$ at $W1<14$, $W1<12.35$,
and $W1<9.7$, respectively, which correspond to members with spectral types
of $\lesssim$M7.5, $\lesssim$M4.5, and $\lesssim$M0.
Most disk-bearing stars have excesses that are at least 1--2~mag larger
than the color thresholds that we have
applied (see Figure~\ref{fig:wise}), so the completeness limits for stars
with excesses in $W3$ and $W4$ are likely fairly high down to even later
spectral types than those estimates ($\sim$M6 and M2).
We have not obtained spectra of all disk-bearing candidates from our survey,
but the number of remaining candidates that lack spectra (30) is small
compared to the total number of known members with disks ($\sim$500), so
they have little effect on our assessment of the completeness of the census 
of disks in Upper Sco.

\subsection{Membership Constraints from Gaia}

\begin{figure}[h]
	\centering
	\includegraphics[trim = 0mm 0mm 0mm 0mm, clip=true, scale=0.8]{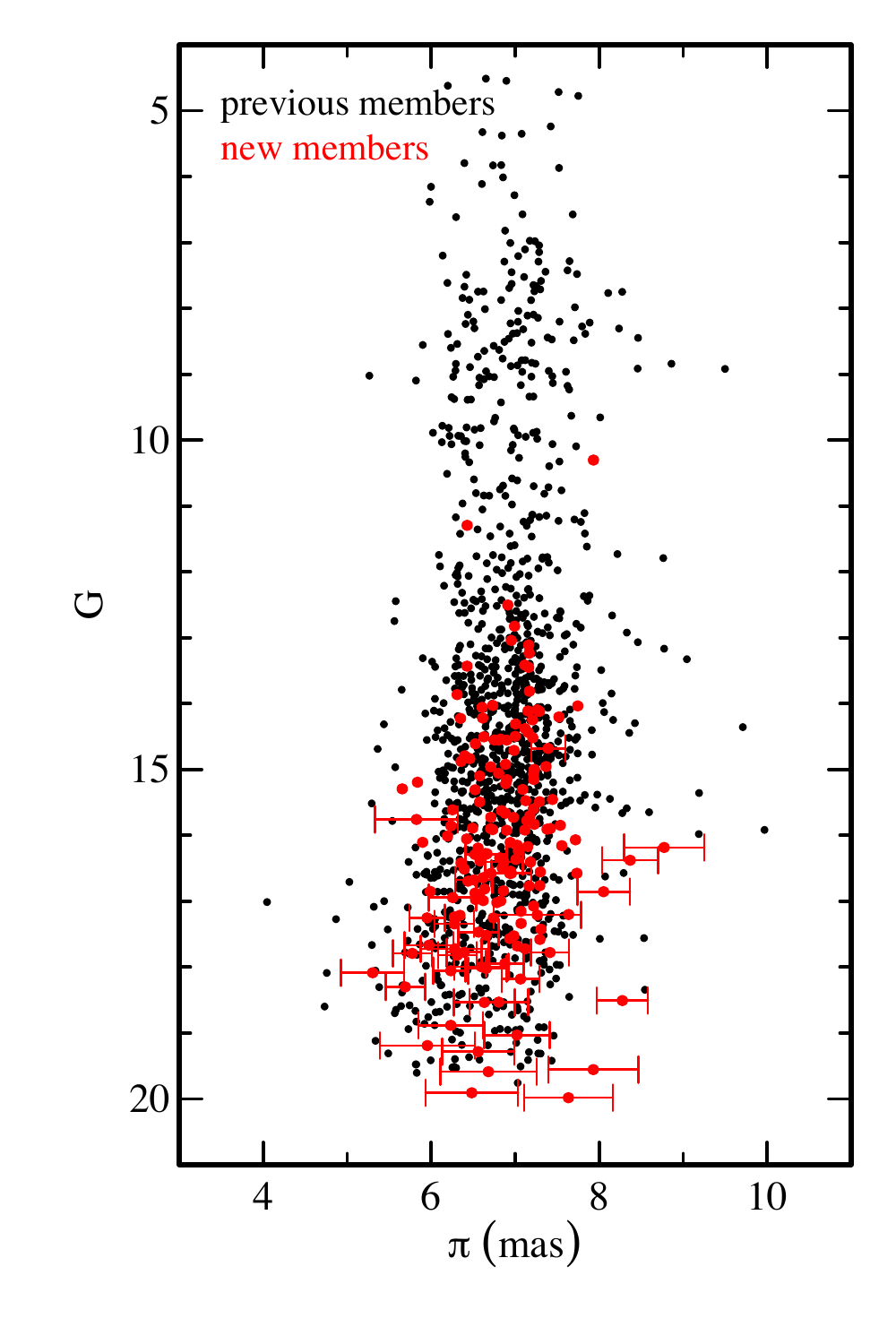}
\caption{
$G$ versus parallax for the previously known members of Upper Sco \citep[black
points,][references therein]{luh18} and the new members from our survey 
(red points) that have proper motions and parallaxes from {\it Gaia} DR2.
Error bars are shown only for new members with errors of $>0.1$~mas.
}
\label{fig:pi}
\end{figure}

\begin{figure}[h]
	\centering
	\includegraphics[trim = 0mm 0mm 0mm 0mm, clip=true, scale=0.8]{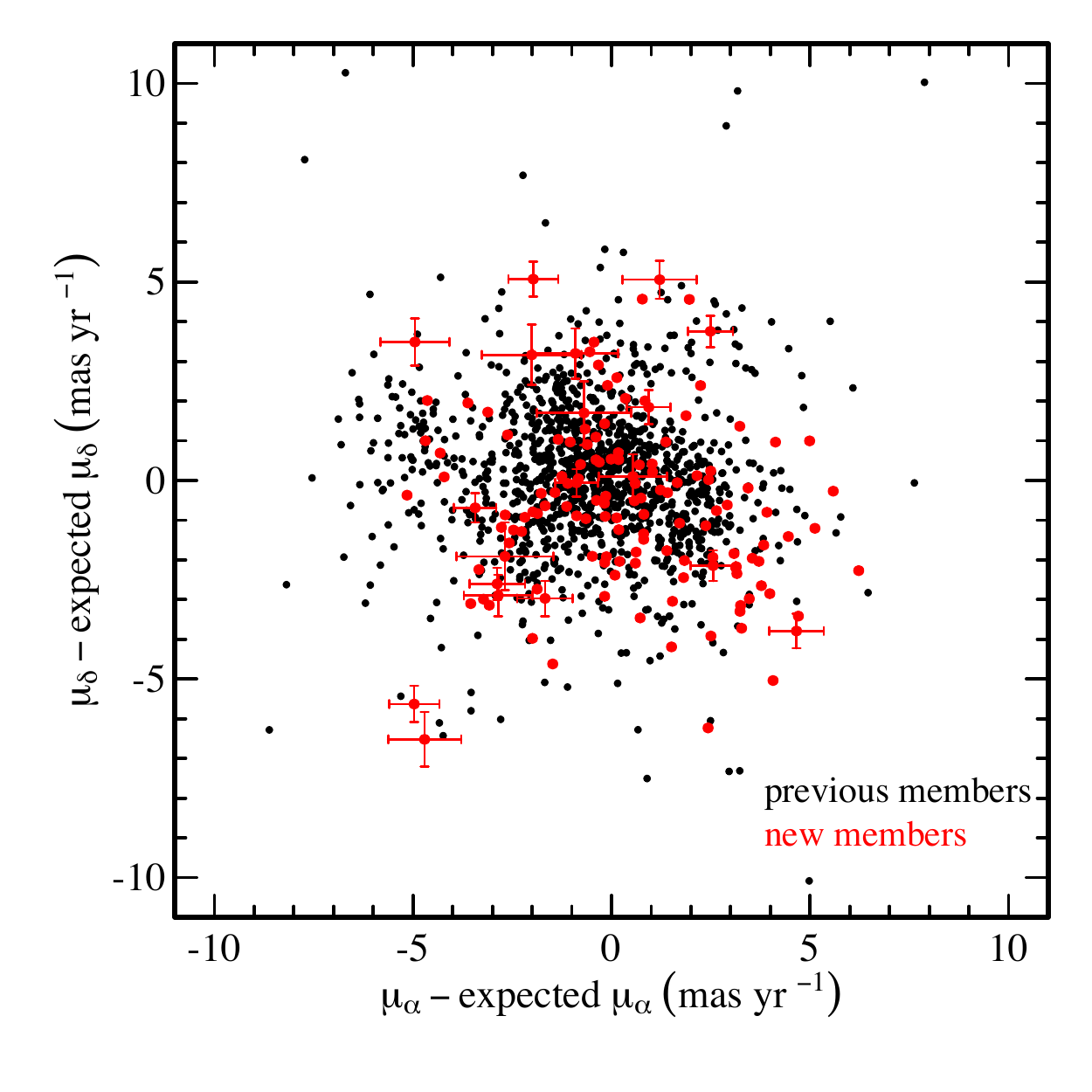}
\caption{
Proper motion offsets for the previously known members of Upper Sco
\citep[black points,][references therein]{luh18} and the new members from our
survey (red points) that have proper motions and parallaxes from {\it Gaia}
DR2. The offsets are relative to the motions expected
for the positions and parallaxes of the stars assuming the median space
velocity of Upper Sco. Errors bars are shown only for new members
with errors of $>0.5$~mas~yr$^{-1}$.
}
\label{fig:pm}
\end{figure}

{\it Gaia} DR2 provides measurements of parallaxes and proper motions for
most of the stars in our spectroscopic sample. 
We can use these data to assess the membership of the spectroscopic targets
that we have classified as members of Upper Sco.
In Figures \ref{fig:pi}, we plot {\it Gaia} $G$ versus parallax for the 1181
previously known members compiled by \citet{luh18} and the 141 new members
from our survey that have {\it Gaia} DR2 parallaxes with errors of $\leq10$\%.
We also compare these two samples in Figure~\ref{fig:pm}, where we 
show the offsets in the {\it Gaia} proper motions relative to the
motions expected for the positions and parallaxes of the stars assuming the
median space velocity of known Upper Sco members that have radial velocity
measurements from {\it Gaia} ($U, V, W=-5.5, -16.3, -6.9$~~km~s$^{-1}$).
In both diagrams, most of the stars that we have classified as new
members have parallaxes and motions similar to those of the previously known
members. Among the new members that are outliers in proper motion or parallax, 
all but one, WISEA J154126.53-261325.5, have large values of the
parameter that characterizes the astrometric goodness of fit, which may
indicate that the astrometry is erroneous, possibly because of a binary
companion. WISEA J154126.53-261325.5 is on the outskirts of Upper Sco and
is an outlier in proper motion but shows strong evidence of youth in 
its spectrum and in the presence of excess emission from a circumstellar disk.
It may be a young interloper, but we adopt it as member of Upper Sco
for purposes of this study.

\section{Disk Classifications}
\label{sec:class}

\subsection{Mid-IR Photometry}

To identify the circumstellar disks in Upper Sco and to classify their
evolutionary stages,
we have compiled all available mid-IR photometry from \spitz\ and \wise\ 
for the catalog of 1631 known members from \cite{luh18},
which includes the new members found in this study.

\subsubsection{\spitz}

The images from \spitz\ were obtained with
the Infrared Array Camera \citep[IRAC;][]{faz04}
and the Multiband Imaging Photometer for Spitzer \citep[MIPS;][]{rie04}. 
IRAC produced images with a field of view of $5\farcm2\times5\farcm2$ and 
FWHM of $1\farcs6$--$1\farcs9$ in four bands centered at 3.6, 4.5, 5.8, and 
8.0~\micron, which are denoted as [3.6], [4.5], [5.8], and [8.0].  
For the band at 24 \micron\ ([24]) utilized in our work, MIPS produced
images with a field of view of $5\farcm4\times5\farcm4$ and a FWHM of
$5\farcs9$.

We have adopted the IRAC and MIPS measurements that are available from
\citet{luh12}. For stars that were observed by \spitz\ but that were absent 
from that study, we have measured aperture photometry from the SEIP mosaics.
The photometry was measured with the {\tt phot} task in IRAF using 
aperture radii of 2.5 pixels for [3.6] and [4.5] and 3 pixels for [5.8],
[8.0], and [24].  The mosaic pixel scales were $0\farcs6$ for IRAC and
$2\farcs45$ for MIPS. The inner/outer radii of the background annuli were
4/7, 4/7, 4/7, 5/7, and 5/7 pixels for the five bands, respectively.
These measurements were calibrated so that the same methods applied to stars
from \citet{luh12} produced photometry that matched the data in that study,
on average. 
For members that were blended with other stars, we applied 
point spread function (PSF) subtraction to remove the contamination
from the neighboring star prior to the aperture photometry.
For each of the four bands of IRAC, the camera detected all members that
were within its field of view with the exception of unresolved companions.
At least one band of IRAC photometry is available for 447 members.
Images from MIPS encompassed 579 members, 424 of which were detected.
Our compilation of IRAC and MIPS photometry for known members
of Upper Sco is presented in Table~\ref{tab:disk}.

\subsubsection{\wise}
\label{sec:wise}

We have retrieved photometry in $W1$--$W4$ for the known members of Upper Sco 
from the \wise\ All-Sky Catalog and the AllWISE Source Catalog.
For bright stars, some data in $W1$ and $W2$ are subject to larger errors
in the latter than the former \citep{cut13}, so we have adopted the data
at $W1<8$ and $W2<7$ from the All-Sky Catalog and otherwise have adopted
the data from the AllWISE Source Catalog.  We have omitted measurements at
$W2<6$ because of their large systematic errors \citep{cut12}.
For the members that lack entries in AllWISE, we inspected
the \wise\ images to check for detections. When detections were present,
we adopted any available measurements from the AllWISE Reject Table and the
WISE All-Sky Source Catalog.
For all members with \wise\ counterparts, we inspected the four bands of
\wise\ images to check for false detections, contamination from diffraction
spikes of other stars, blending with other stars, extended emission,
and offsets in the centroids from shorter to longer wavelengths.
The latter can indicate that a very red source (typically a galaxy) is blended
with the target and is dominant in $W3$ and $W4$.
We excluded photometry of this kind if a contaminating source could be
identified in other images with higher resolution (e.g., UKIDSS).
Otherwise, if a contaminant could not be identified in other data, the
presence of an offset centroid is noted in Table~\ref{tab:disk}.

Our compilation of \wise\ data is included with the \spitz\ photometry
in Table~\ref{tab:disk}. 
At least one band of \wise\ photometry is available for 1598 members.
The members that lack \wise\ photometry are unresolved from
brighter stars, saturated, or not detected.
Table~\ref{tab:disk} contains a total of 1608 members that have photometry
from either \spitz\ or \wise. The remaining 23 known members that lack
data from either facility are not included in Table~\ref{tab:disk}.
These objects are too faint for detection, unresolved from brighter stars
or diffraction spikes, or saturated in all bands (Antares).

\subsection{Measurements of Excess Emission}
\label{sec:excess}

As done in a similar study of IRAC and {\it WISE} data in Upper Sco
by \cite{luh12}, we have searched for the presence of IR excess emission
from disks among the known members of the association using colors between
$K_s$ and [4.5], [8.0], [24], $W2$, $W3$, and $W4$.
We have included our adopted values of $K_s$ in Table~\ref{tab:disk}.
When computing the colors, we have ignored data with errors $>0.1$ and $>0.25$ 
for [4.5]/[8.0] and [24]/$W3$/$W4$, respectively.
We have corrected the colors for extinction using the estimates of $A_K$
from \citet{luh18} and the following extinction relations:
$A_{4.5}/A_K=0.5$, $A_{8.0}/A_K=0.45$, $A_{24}/A_K=0.3$,
$A_{W2}/A_K=0.48$, $A_{W3}/A_K=0.55$, and $A_{W4}/A_K=0.4$
\citep[][references therein]{asc13,sch16,xue16}.

In Figure~\ref{fig:excess}, we plot the extinction-corrected IR colors of
the known members of Upper Sco as a function of spectral type.
As in the \wise\ CMDs and color-color diagrams from Section~\ref{sec:search},
we have marked the stars with disks (Section~\ref{sec:term}).
In each color, the data exhibit a well-defined sequence of blue colors
that correspond to stellar photospheres and a wide range of redder colors
that likely indicate the presence of excess emission from disks.
To identify the stars that have color excesses,
we began by deriving a fit to the photospheric sequence in each color. 
For the latest spectral types where some of the sequences are not
well-populated, we adopted the colors of young photospheres measured
in other star-forming regions \citep[][K. Luhman in preparation]{luh10}.
We did not attempt to identify excesses for objects later than L0 since
their intrinsic colors are uncertain.
In each color, we defined a boundary that follows the lower envelope
photospheric sequence and used the reflection of that boundary above the
sequence as a threshold for identifying excesses, as indicated in
Figure~\ref{fig:excess}. Similar criteria for excesses were applied
to \spitz\ and \wise\ data in Upper Sco and Taurus by \citet{luh12}
and \citet{esp14}.

For \wise\ sources that have offset centroids in $W3$ or $W4$
(Section~\ref{sec:wise}, ``offset" in Table~\ref{tab:disk}), we report 
an excess in those bands only if an excess is also present in bands at
shorter wavelengths that do not show a shifted centroid.
WISEA J160600.62$-$195711.8 is a $1\arcsec$ pair (based on UKIDSS images)
that lacks resolved
spectral types for its components and is not resolved by \wise, so it 
appears as a single entry in Table~\ref{tab:disk}. The $W4$ centroid
for the pair appears to be centered on the southern component, which is
therefore responsible for the $W4$ excess.
If a given band exhibited an excess but data at longer wavelengths were
inconsistent with an excess, that band was not marked as showing an excess.
The presence of an excess is marked as unknown or uncertain for some stars
because of contamination by diffraction spikes, extended emission, or nearby
stars. If only a marginal excess was present in a band and data at longer
wavelengths were unavailable, the excess was marked as tentative.
Among the 1608 members of Upper Sco that have photometry
from either \spitz\ or \wise\ (Table~\ref{tab:disk}), we find excesses
and tentative excesses in at least one band for 489 and 34 sources,
respectively.

\subsection{Disk Classes in Upper Sco}
\label{sec:term}

As done in our previous studies of disks in Upper Sco and Taurus
\citep{luh12,esp14}, we have classified the evolutionary 
status of the circumstellar disks in Upper Sco using 
a scheme that includes the following classes
\citep{ken05,rie05,her07,luh10,esp12}:
{\it full disks} are optically thick at IR wavelengths and lack
significant clearing of primordial dust and gas;
{\it pre-transitional disks} and {\it transitional disks} have large inner
gaps or holes in their dust distributions, respectively;
{\it evolved disks} are becoming optically thin but have not undergone
significant clearing;
{\it evolved transitional disks} are optically thin and have large holes;
{\it debris disks} consist of dust produced by collisions of planetesimals.

\begin{figure}[h]
	\centering
	\includegraphics[trim = 0mm 0mm 0mm 0mm, clip=true, scale=1]{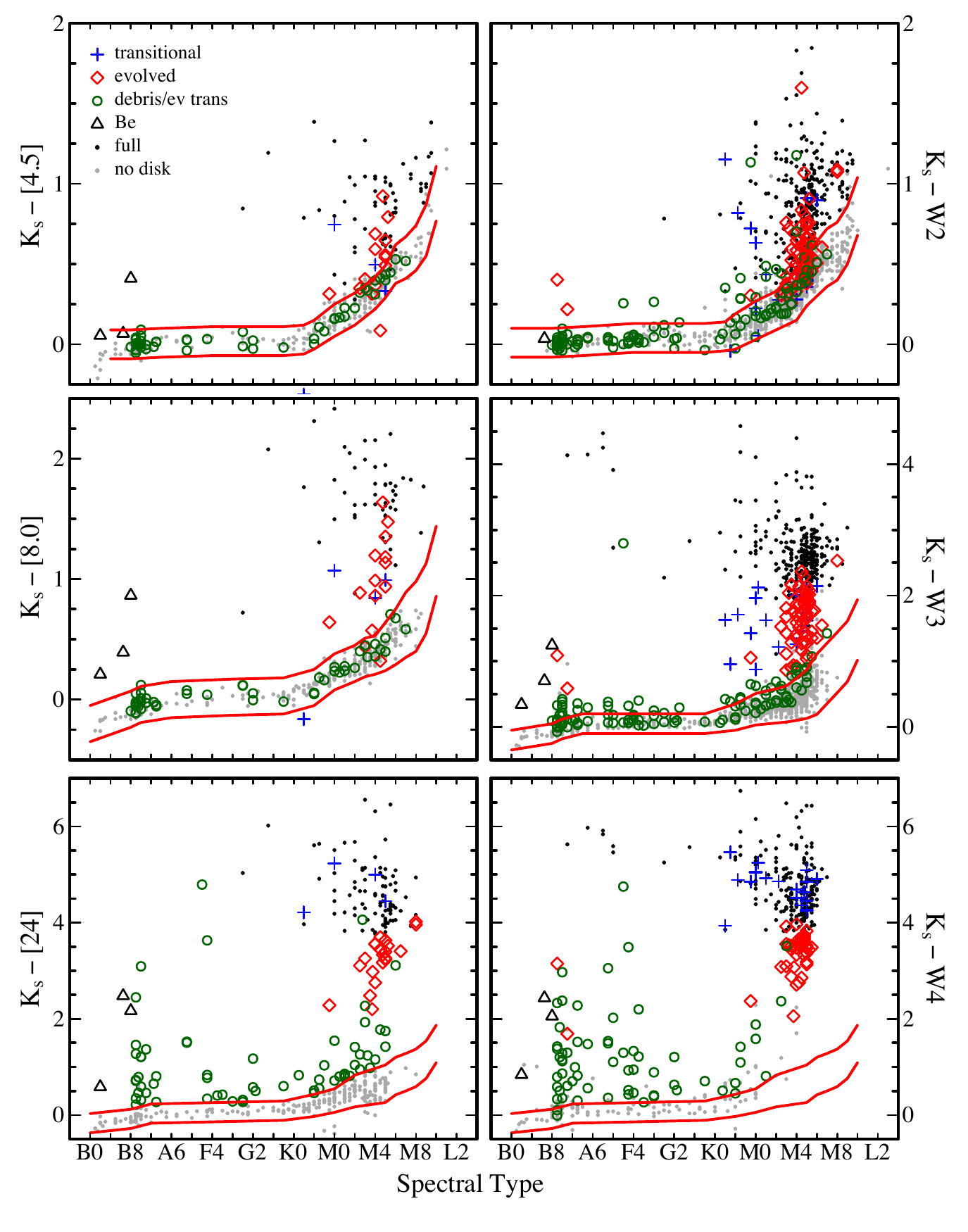}
\caption{
Extinction-corrected IR colors as a function of spectral type for the members
of Upper Sco using data from {\it Spitzer} (left) and {\it WISE} (right)
with $K_s$ data from 2MASS and UKIDSS. 
In each color, we have selected a boundary that follows the lower envelope
the members and we have marked the reflection of that boundary above the
blue sequence (solid lines). Those upper boundaries are used to identify
the presence of color excesses from circumstellar disks (Table \ref{tab:disk}).
}
\label{fig:excess}
\end{figure}

\begin{figure}[h]
	\centering
	\includegraphics[trim = 0mm 0mm 0mm 0mm, clip=true, scale=1]{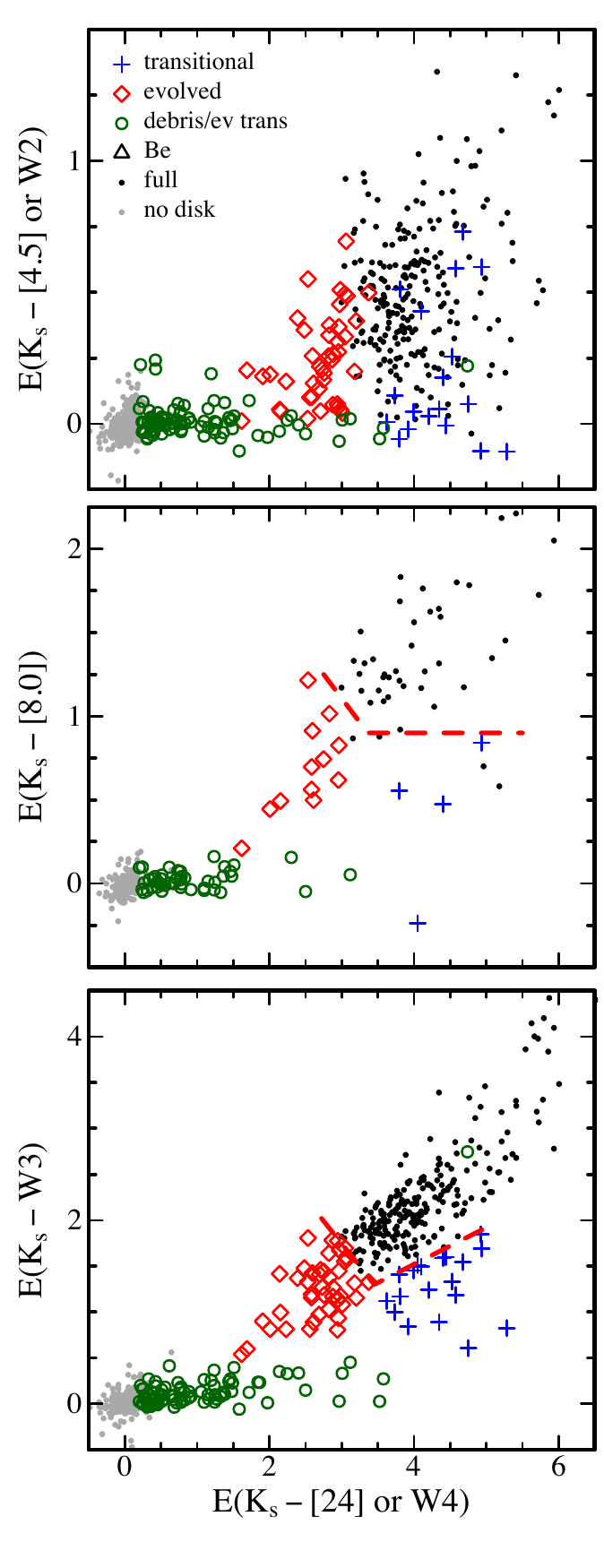}
\caption{
Extinction-corrected IR color excesses for members of Upper Sco. 
Data at [4.5] and [24] are shown when available. 
Otherwise, data from the similar bands of $W2$ and $W4$ are used.
In the bottom two diagrams, we indicate the boundaries that are used to 
distinguish full disks from disks in more advanced stages of evolution.}
\label{fig:diskclass}
\end{figure}

For members of Upper Sco found to exhibit IR excesses in the previous
section, we have estimated disk classes using extinction-corrected color 
excesses (e.g., $E(K_s-[8.0])$), which are computed using the fits to the 
photospheric sequences in Figure~\ref{fig:excess} that we previously described.
The resulting excesses at [4.5], [8.0], $W3$, and [24] are plotted in 
Figure~\ref{fig:diskclass}.  For sources that lack [4.5] or [24], we show 
the excesses in $W2$ or $W4$.
To distinguish full disks from disks that are more evolved, 
we have applied the color excess thresholds defined by \cite{esp14}
for $E(K_s-[8.0])$, $E(K_s-W3)$, and $E(K_s-[24]/W4)$,
which are indicated in Figure~\ref{fig:diskclass}.  
We have modified the threshold in the diagram with $W3$ and [24]/$W4$ so
that it extends to ($E(K_s-[24]/W4),E(K_s-W3)) = (5.0, 1.92)$.
Sources above at least one of the two boundaries shown in 
Figure~\ref{fig:diskclass} are classified as full disks.
Pre-transitional disks are encompassed by these criteria for full disks.
Among the remaining objects with bluer color excesses, we applied
the same criteria for transitional, evolved, evolved transitional, and
debris disks as in \citet{esp14} except that we have revised the threshold
for $E(K_s-[24]/W4)$ from 3.55 to 3.6 to slightly better align with a natural
separation between transitional and evolved transitional/debris disks
(see Figure~\ref{fig:excess}).
Although WISEA J161055.09$-$253121.9 (HIP 79288) is above the full disk
boundaries, it is treated as a debris or evolved transitional disk 
as in previous studies \citep[][references therein]{luh12}.
For stars that do not appear in Figure~\ref{fig:excess} because they
lack photometry in one of the necessary bands for that diagram (particularly
[24] and $W4$), we have estimated the disk classes in the same manner as
in \citet{luh12} and \citet{esp14}.

Our disk classifications are included in Table~\ref{tab:disk}.
Among the 484 members with excesses in at least one band 
(excluding five Be stars),
we classify 296 disks as full, 66 as evolved, 19 as transitional, 
22 as evolved or transitional, and 81 as evolved transitional or debris.
Many of these disks have not been previously reported,
including 129 full disks and 50 disks that are at more advanced 
evolutionary stages.
In addition, we find tentative evidence for one evolved disk, 
three full disks, and 30 debris or evolved transitional disks.
As mentioned in Section~\ref{sec:subsearch}, one member of Upper Sco,
WISEA J161935.70$-$195043.0, has sufficiently red mid-IR colors that
it appears to be an edge-on full disk or a class~I protostar.

\subsection{Comparison to Previous Classifications}

For disks in Upper Sco that have been previously classified, most of
our classifications agree with those from the previous studies 
\citep[e.g.,][]{luh12,daw13,riz15,pec16}.
We discuss a few exceptions in this section.

\cite{cod18} identified a sample of 288 stars toward Upper Sco and Ophiuchus
that exhibit mid-IR excess emission from disks and that have well-measured
light curves from {\it Kepler}'s K2 mission.
Among the 190 stars from that sample that are within our survey field
for Upper Sco, 179 are in the compilation of known members that
we have examined for disks \citep{luh18}.
We do not find IR excesses for three of those 179 stars, which consist of
EPIC 210282528, EPIC 203385048, and EPIC 204637622 (WISEA J163334.90$-$183254.4,
WISEA J161816.17$-$261908.3, WISEA J160420.97$-$213041.6).
An absence of disks is consistent with the K2 variability of those stars
according to the analysis of \citet{cod18}.
Among the 11 stars from \citet{cod18} that are absent from the list of members
that we considered, six are in our sample of remaining disk-bearing candidates
that lack spectroscopy (EPIC 203716389, EPIC 203789325, EPIC 204347824,
EPIC 204397879, EPIC 204487447, EPIC 204982702)
and five did not satisfy the criteria we applied to the \wise\ data
(EPIC 204495624, EPIC 204874314, EPIC 203604427, EPIC 203337814,
EPIC 203082998).  Among the latter five stars, EPIC 204495624 does
show excesses in $W3$ and $W4$ but it was not selected as a candidate because
the errors in those bands were too high for consideration.
EPIC 204874314 has a small $W4$ excess that is below our threshold.
The other three do not appear to show excesses when unreliable
\wise\ detections are excluded.

\cite{lod18} reported possible excesses in $K-W2$ 
for two L-type members, UGCS J160731.61$-$214654.4 (VISTA 1607$-$2146) and 
WISEA J161144.37$-$221544.6 (VISTA J1611$-$2215). 
In images from UKIDSS, the former is $\sim$$3\farcs7$ from a $1\farcs6$ pair
of objects. These three UKIDSS sources are blended and appear as a single
source in the images from \wise. The centroid of the \wise\ detection is
roughly midway between the L-type object and the pair, indicating that the pair
contribute significantly to the flux. As a result, we conclude that the
\wise\ data cannot be used to assess the presence of a disk.
For the other L-type member from \citet{lod18} with a possible excess,
WISEA J161144.37$-$221544.6, the intrinsic photospheric colors near its
spectral type (L3) are too uncertain to determine whether an excess is present.
Its value of $K_s-W2$ is only slightly higher than our adopted threshold
for L0 (see Figure~\ref{fig:excess}). If the intrinsic colors of young
L-type objects continue to become redder beyond L0, then its $K_s-W2$
would not exhibit an excess.

\section{Conclusion}

We have performed a survey for new disk-bearing members of Upper Sco
and have identified and classified the disks among all known members of the
association. Our results are summarized as follows:

\begin{enumerate}
\item We have identified candidate disk-bearing members of Upper Sco 
via mid-IR excesses in photometry from \wise\ and have further refined
the resulting sample with photometry from \spitz, 2MASS, and USNO
and proper motions from UCAC4 and UCAC5.
Through optical and near-IR spectroscopy, we have confirmed 185
candidates as likely members of Upper Sco (mid-K through M9),
19 of which have been reported in previous studies.
These 185 members comprise $\sim$36\% of the known disk-bearing stars in
the association. We also have found 29 new stars with disks in the outskirts 
of Ophiuchus.

\item We have compiled all available mid-IR photometry from
\spitz\ and \wise\ for the known members of Upper Sco adopted by
\citet{luh18}, which include those found in this study.
We find that 1608 members have photometry from at least
one of the two facilities. Our catalog serves as an update to a
previous compilation of \spitz\ and \wise\ data for members by \citet{luh12}.
Through that study and the one that we present,
all \spitz\ and \wise\ images of the known members have been visually 
inspected to check for false detections and contamination by diffraction 
spikes, extended emission, and nearby stars.
 
\item 
We have searched for the presence of IR excess emission from disks among 
the known members of Upper Sco using several colors between $K_s$ and bands
from \spitz\ and \wise. For the 484 stars that exhibit significant
excesses (excluding five Be stars), we have classified the evolutionary stages
of the disks \citep{esp12} with criteria similar to those applied in our 
previous studies of Taurus and Upper Sco \citep{luh12,esp14}.
We classify 296 disks as full, 66 as evolved, 19 as transitional,
22 as evolved or transitional, and 81 as evolved transitional or debris.
Many of these disks have not been previously reported, including 129 full 
disks and 50 disks that are at more advanced evolutionary stages.

\end{enumerate}

\acknowledgements

T.E., K.L., and E.B. were supported by grant NNX12AI58G from the NASA 
Astrophysics Data Analysis Program. E.M. acknowledges support from the NASA 
NExSS program.  We thank Katelyn Allers for providing the modified version of
Spextool for use with ARCoIRIS data.
\wise\ and {\it NEOWISE} are joint projects of
the University of California, Los Angeles, and the Jet Propulsion
Laboratory (JPL)/California Institute of Technology (Caltech), funded by NASA.
The {\it Spitzer Space Telescope} is operated by JPL and Caltech under
contract with NASA.  The IRTF is operated by the University of Hawaii under
contract NNH14CK55B with NASA.
2MASS is a joint project of the University of
Massachusetts and the Infrared Processing and Analysis Center (IPAC) at
Caltech, funded by NASA and the NSF.
Our work is based in part on (1) observations at CTIO, National Optical Astronomy Observatory (NOAO Prop. ID: 2015A-0192 and 2017A-0161; PI: T. Esplin), which is operated by the Association of Universities for Research in Astronomy (AURA) under a cooperative agreement with the NSF;
and (2) observations obtained at the SOAR telescope, which is a joint project of the Minist\'{e}rio da Ci\^{e}ncia, Tecnologia, Inova\c{c}\~{a}os e Comunica\c{c}\~{a}oes (MCTIC) do Brasil, the NOAO, the University of North Carolina at Chapel Hill (UNC), and Michigan State University (MSU).
The Digitized Sky Surveys were produced at the Space Telescope Science Institute under U.S. Government grant NAG W-2166. The images of these surveys are based on photographic data obtained using the Oschin Schmidt Telescope on Palomar Mountain and the UK Schmidt Telescope. The plates were processed into the present compressed digital form with the permission of these institutions.
This work has made use of data from the European Space Agency (ESA)
mission {\it Gaia} (\url{https://www.cosmos.esa.int/gaia}), processed by
the {\it Gaia} Data Processing and Analysis Consortium (DPAC,
\url{https://www.cosmos.esa.int/web/gaia/dpac/consortium}). Funding
for the DPAC has been provided by national institutions, in particular
the institutions participating in the {\it Gaia} Multilateral Agreement.
The Center for Exoplanets and Habitable Worlds is supported by the
Pennsylvania State University, the Eberly College of Science, and the
Pennsylvania Space Grant Consortium.

{\it Facilities: } \facility{Blanco (COSMOS, ARCoIRIS)}, 
\facility{SOAR (Goodman)},
\facility{IRTF (SpeX)},
\facility{WISE},
\facility{Spitzer}

\clearpage

\begin{deluxetable}{cc}
\tabletypesize{\scriptsize}
\tablewidth{0pt}
\tablecaption{Vertices of Adopted Boundary between Upper Sco and Ophiuchus\label{tab:oph}}
\tablehead{
\colhead{Right Ascension (J2000)} & \colhead{Declination (J2000)} \\
\colhead{(deg)} &
\colhead{(deg)}}
\startdata
246.157 & $-$24.60 \\
245.250 & $-$23.40 \\
245.750 & $-$22.75 \\
246.750 & $-$22.75 \\
246.750 & $-$23.80 \\
247.510 & $-$23.80 \\
248.200 & $-$24.25 \\
248.200 & $-$24.25 \\
249.300 & $-$25.50 \\
248.100 & $-$25.00 \\
247.510 & $-$25.00 \\
246.100 & $-$25.00 
\enddata
\end{deluxetable}

\clearpage

\begin{deluxetable}{llll}
\tabletypesize{\scriptsize}
\tablewidth{0pt}
\tablecaption{Observing Log\label{tab:log}}
\tablehead{
\colhead{Telescope/Instrument} &
\colhead{Disperser/Aperture} &
\colhead{Wavelengths/Resolution} &
\colhead{Targets}}
\startdata
CTIO 4~m/COSMOS & red VPH/$0\farcs9$ slit & 0.55--0.95~\micron/3~\AA & 101 \\
SOAR/Goodman & 400 l~mm$^{-1}$/$0\farcs84$ slit & 0.54-0.94~\micron/6~\AA  & 34 \\
CTIO 4~m/ARCoIRIS & $1\farcs1$ slit & 0.8--2.47~\micron/R=3500 & 10 \\
IRTF/SpeX & prism/$0\farcs8$ slit & 0.8--2.5~\micron/R=150 & 89
\enddata
\end{deluxetable}

\clearpage

\begin{deluxetable}{lcccccc}
\tabletypesize{\scriptsize}
\tablewidth{0pt}
\tablecaption{Spectroscopic Data for Candidate Members of Upper Sco\label{tab:specu}}
\tablehead{
\colhead{Source Name\tablenotemark{a}} &
\colhead{Spectral Type} & 
\colhead{$W_\lambda$(Li)} &
\colhead{$W_\lambda$(Na)} &
\colhead{Instrument} &
\colhead{Date} &
\colhead{Member?}\\
\colhead{} &
\colhead{} &
\colhead{(\AA)} &
\colhead{(\AA)} &
\colhead{} &
\colhead{} &
\colhead{}}
\startdata
WISEA J154126.53-261325.5 & M5.75 & \nodata & \nodata & SpeX & 2015 Apr 21 & Y \\
WISEA J154419.28-231306.9 & M6 & \nodata & 3.2 & COSMOS & 2015 May 11 & Y \\
WISEA J154425.49-212641.0 & M3.25 & 0.45 & 2.6 & Goodman & 2014 Jun 16 & Y \\
WISEA J154626.94-244322.8 & M4.75 & 0.70 & 3.7 & COSMOS & 2015 May 11 & Y \\
WISEA J154824.44-223549.8 & M3.75 & 0.70 & 3.0 & COSMOS & 2015 May 13 & Y
\enddata
\tablecomments{This table is available in its entirety in a machine-readable form. A portion is shown is here for guidance regarding its form and content.}
\tablenotetext{a}{Coordinate-based identifications from the AllWISE Source
Catalog when available. Otherwise, identifications are from the WISE All-Sky
Source Catalog.}
\end{deluxetable}

\clearpage

\begin{deluxetable}{lcccccc}
\tabletypesize{\scriptsize}
\tablewidth{0pt}
\tablecaption{Spectroscopic Data for Candidate Members of Ophiuchus\label{tab:speco}}
\tablehead{
\colhead{Source Name\tablenotemark{a}} &
\colhead{Spectral Type} & 
\colhead{$W_\lambda$(Li)} &
\colhead{$W_\lambda$(Na)} &
\colhead{Instrument} &
\colhead{Date} &
\colhead{Member?}\\
\colhead{} &
\colhead{} &
\colhead{(\AA)} &
\colhead{(\AA)} &
\colhead{} &
\colhead{} &
\colhead{}}
\startdata
WISEA J162224.95-232955.1 & M3.5 & 0.60 & 2.3 & COSMOS & 2015 May 13 & Y \\
WISEA J162235.61-233733.9 & M2 & \nodata & \nodata & ARCoIRIS & 2016 Jun 19 & Y \\
WISEA J162247.18-230013.4 & M3.5 & \nodata & \nodata & ARCoIRIS & 2016 Jun 19 & Y \\
WISEA J162248.63-230218.0 & M5.75 & \nodata & \nodata & SpeX & 2015 Apr 19 & Y \\
WISEA J162308.78-225743.0 & M0.25 & 0.50 & 2.5 & Goodman & 2014 Jun 17 & Y 
\enddata
\tablecomments{This table is available in its entirety in a machine-readable form. A portion is shown is here for guidance regarding its form and content.}
\tablenotetext{a}{Coordinate-based identifications from the AllWISE Source
Catalog.}
\end{deluxetable}

\clearpage

\begin{deluxetable}{lrrrr}
\tabletypesize{\scriptsize}
\tablewidth{0pt}
\tablecaption{Candidate Disk-bearing Members of Upper Sco that Lack 
Spectroscopy\label{tab:cand}}
\tablehead{
\colhead{Source Name\tablenotemark{a}} &
\colhead{$W1$} & 
\colhead{$W2$} & 
\colhead{$W3$} & 
\colhead{$W4$}}
\startdata
WISEA J154255.15$-$253610.2 &  9.77$\pm$0.02 &  9.64$\pm$0.02 &  9.05$\pm$0.04 & 7.75$\pm$0.19 \\
WISEA J154703.73$-$260118.6 & 12.05$\pm$0.02 & 11.76$\pm$0.02 & 10.46$\pm$0.10 & 8.40$\pm$0.34 \\
WISEA J154919.75$-$225729.8 &  9.39$\pm$0.02 &  9.17$\pm$0.02 &  8.68$\pm$0.03 & 6.91$\pm$0.09 \\
WISEA J155125.62$-$270743.4 & 11.33$\pm$0.02 & 11.16$\pm$0.02 & 10.07$\pm$0.06 & 7.83$\pm$0.24 \\
WISEA J155233.92$-$265112.5 & 11.04$\pm$0.02 & 10.53$\pm$0.02 &  9.29$\pm$0.04 & 8.52$\pm$0.39 
\enddata
\tablecomments{This table is available in its entirety in a machine-readable form. A portion is shown is here for guidance regarding its form and content.}
\tablenotetext{a}{Coordinate-based identifications from the AllWISE Source
Catalog.}
\end{deluxetable}
\clearpage

\begin{deluxetable}{ll}
\tabletypesize{\scriptsize}
\tablewidth{300pt}
\tablecaption{Mid-IR Photometry for Members of Upper Sco\label{tab:disk}}
\tablehead{
\colhead{Column Label} &
\colhead{Description}}
\startdata
Name & Source name\tablenotemark{a} \\
RAdeg & Right Ascension (J2000) \\
DEdeg & Declination (J2000) \\
SpType & Adopted spectral type\tablenotemark{b} \\
Ksmag & $K_s$ band magnitude \\
e\_Ksmag & Error in $K_s$ band magnitude \\
r\_Ksmag & Reference for $K_s$ band magnitude\tablenotemark{c}  \\
Ak & Extinction in $K_s$\tablenotemark{b} \\
3.6mag & {\it Spitzer} [3.6] band magnitude \\
e\_3.6mag & Error in [3.6] band magnitude \\
f\_3.6mag & Flag on [3.6] band magnitude\tablenotemark{d} \\
4.5mag & {\it Spitzer} [4.5] band magnitude \\
e\_4.5mag & Error in [4.5] band magnitude \\
f\_4.5mag & Flag on [4.5] band magnitude\tablenotemark{d} \\
5.8mag & {\it Spitzer} [5.8] band magnitude \\
e\_5.8mag & Error in [5.8] band magnitude \\
f\_5.8mag & Flag on [5.8] band magnitude\tablenotemark{d} \\
8.0mag & {\it Spitzer} [8.0] band magnitude \\
e\_8.0mag & Error in [8.0] band magnitude \\
f\_8.0mag & Flag on [8.0] band magnitude\tablenotemark{d} \\
24mag & {\it Spitzer} [24] band magnitude \\
e\_24mag & Error in [24] band magnitude \\
f\_24mag & Flag on [24] band magnitude\tablenotemark{d} \\
W1mag & {\it WISE} $W1$ band magnitude\tablenotemark{e} \\
e\_W1mag & Error in $W1$ band magnitude \\
f\_W1mag & Flag on $W1$ band magnitude\tablenotemark{d} \\
W2mag & {\it WISE} $W2$ band magnitude\tablenotemark{e} \\
e\_W2mag & Error in $W2$ band magnitude \\
f\_W2mag & Flag on $W2$ band magnitude\tablenotemark{d} \\
W3mag & {\it WISE} $W3$ band magnitude\tablenotemark{e} \\
e\_W3mag & Error in $W3$ band magnitude \\
f\_W3mag & Flag on $W3$ band magnitude\tablenotemark{d} \\
W4mag & {\it WISE} $W4$ band magnitude\tablenotemark{e} \\
e\_W4mag & Error in $W4$ band magnitude \\
f\_W4mag & Flag on $W4$ band magnitude\tablenotemark{d} \\
Exc4.5 & Excess present in [4.5]? \\
Exc8.0 & Excess present in [8.0]? \\
Exc24 & Excess present in [24]? \\
ExcW2 & Excess present in $W2$? \\
ExcW3 & Excess present in $W3$? \\
ExcW4 & Excess present in $W4$? \\
DiskType & Disk Type
\enddata
\tablecomments{This table is available in its entirety in a machine-readable form.}
\tablenotetext{a}{Coordinate-based identifications from the AllWISE Source
          Catalog when available. Otherwise, identifications are from
          the WISE All-Sky Source Catalog, 2MASS All-Sky Point Source
          Catalog, or Data Release 10 of the UKIDSS Galactic Clusters Survey.}
\tablenotetext{b}{\citet{luh18}.}
\tablenotetext{c}{2 = 2MASS Point Source Catalog; u = UKIDSS Data Release 10}
\tablenotetext{d}{nodet = non-detection; sat = saturated; out = outside of the
          camera's field of view; bl = photometry may be affected by
          blending with a nearby star; ext = photometry is known or
          suspected to be contaminated by extended emission (no data
          given when extended emission dominates); dif = photomtery
          may be affected by a diffraction spike; bin = includes an
          unresolved binary companion; unres = too close to a brighter
          star to be detected; false = detection from WISE catalog
          appears false or unreliable based on visual inspection;
          off = $W3$ and/or $W4$ detection appears offset from the $W1/W2$
          detection (no data given when offset is due to a known
          source); err = $W2$ magnitudes brighter than $\sim$6 are 
          erroneous.}
\tablenotetext{e}{Photometry from AllWISE Source Catalog except for
          WISEAR J161336.90-232730.0, WISEAR J163810.80-294040.6,
          and WISEAR J160017.31-223650.8, which are from the
          AllWISE Reject Table, and WISE J160414.16-212915.5,
          WISE J160027.15-223850.5, WISE J161320.78-175752.3,
          WISE J162230.38-241119.2, WISE J162210.14-240905.4,
          WISE J161317.38-292220.0, WISE J161837.22-240522.8, and
          WISE J162620.15-223312.8, which are from the WISE
          All-Sky Source Catalog.}
\end{deluxetable}

\clearpage

\end{document}